\documentclass[aps,amsmath,twocolumn]{revtex4}
\usepackage{graphicx}
\usepackage{subfig}
\usepackage{relsize}

\begin{document}


\title{A new kind of radio transient: ERBs}

\author{Douglas Scott} \email{docslugtoast@phas.ubc.ca}
\author{Ali Frolop} \email{afrolop@phas.ubc.ca}
\affiliation{Dept.\ of Physics \& Astronomy,
 University of British Columbia, Vancouver, Canada}

\date{1st April 2019}

\begin{abstract}
We describe the discovery of a new kind of radio transient, which we call
``early-riser bursts'' or ERBs.  We found this new class of source by
considering traditional radio searches, but extending into the complex
plane of dispersion measure.  ERBs have the remarkable property of
appearing before they are searched for.  We provide suggestions for the
most likely origin of this new astronomical phenomenon.
\end{abstract}

\maketitle

\noindent
\section{Introduction}
Time-domain astronomy is exploding.  It used to be that progress
happened in a uniform and continuous way, but increasingly in the past few
decades we have seen abrupt eruptions of activity, with new science ideas
cropping up unexpectedly.

An obvious example of this is the sudden appearance of models of transient
events in the radio sky \cite{dynamic}.
Some of these ideas have even been accompanied by
empirical evidence for the existence of such events, e.g.\ ``fast radio
bursts'' or FRBs.

Originally discovered using data from the Parkes Observatory, FRBs are now
being routinely detected using the CHIME instrument \cite{chime}.
This year, for the first time there are probably more individual FRB events
than different
theories to explain them \cite{theories}, and so it is time to consider
whether there may be entirely new classes of transient source to discover.

\section{Detection of FRBs}

Knowing that FRBs were originally found by broadening the parameter
space in searches for pulsars, we want to further extend how transient events
are found in radio data to see if we can uncover something new.

A key phenomenon to consider for the detection of FRBs is that the
radio waves are dispersed by the plasma through which they travel,
so that high frequencies arrive earlier than low frequencies.
The time delay depends on a quantity called ``dispersion
measure'', usually given the variable name ``DM'' \cite{DM}.  The delay between
frequencies $\nu_{\rm low}$ and $\nu_{\rm high}$ is then
\begin{equation}
{\Delta t} \propto
 \left({1\over \nu_{\rm low}} - {1\over \nu_{\rm high}}\right) \times {\rm DM}.
\end{equation}
One has to correct for this effect in order to optimise the signal-to-noise
ratio of the burst.  However, FRBs are detected with a wide set of values of
dispersion measure and hence it is necessary to search over a range of DM
when looking for transient signals that are dispersed according to frequency.

Several algorithms exist for performing this search task, e.g.\
direct dedispersion \cite{directde},
brute-force dedispersion \cite{brutede},
tree dedispersion \cite{treede},
subband dedispersion \cite{subbandde},
subversion dedispersion \cite{svnde}
and cast-aspersion dedispersion \cite{cade}.
We have developed a new approach
for undoing the effects of dispersion, which we just call ``{\tt spersion}''.
We then used this method to search for new kinds of radio
transient.

It is well known that the first FRB, the Lorimer Event, was discovered when
a search was mounted over a wider range of DM than normally considered
\cite{lorimer}.
Inspired by this, we decided to search the radio archive for sources with
{\it negative\/} DM values.  This would be a natural signal for alien
intelligences to send to us, since it is not something that the interstellar
medium can make naturally.  Inspired by recent suggestions
\cite{seti} that unusual
astronomical signals might have their origin in extraterrestrial civilisations
\cite{ets}, we used {\tt spersion} to perform a search into the
negative DM region.

Unfortunately this investigation was unsuccessful,
and we found no obvious evidence
of bursts of this kind produced by intelligent aliens.  However, undeterred,
we then extended our net further and found something even more remarkable.

\section{Detection of ERBs}

Most dedispersion methods use intensity data, i.e.\ the incoherent signal,
coming from the square of the electric fields.  It is possible instead to
directly examine the coherent data.  Dedispersion then corresponds to
multiplying the $E$ field signal by a (frequency-dependent) pure phase.
By extending the phase factors into the full complex plane we can broaden the
scope of dedispersion patterns that are applied to interferometric signals.
We specifically found that it is interesting to explore
imaginary DMs, in analogy with the use of ``imaginary time'' in certain
approaches to special relativity and quantum mechanics, i.e.\ we are
considering  a Wick rotation in the DM complex plane \cite{wick}.

Carefully searching through the data archive in this way, we discovered
the first example of an entirely new class of source.  We call these
``early-riser bursts'' or ERBs.  This kind of event is even more
unexpected than we expected.  Because of the complex dedispersion applied
to the finite-time window of the radio data sets, these ERBs
literally happen {\it just before\/} you look for them.  This behaviour
explains why they had previously escaped attention.  Such a
phenomenon is not unknown to science,
with the ERBs behaving in a similar way to the properties of ``thiotimoline'',
described more than half a century ago in chemistry \cite{thiotimoline}.

\section{What are ERBs?}

There are several possibilities for the origin of these early-riser bursts,
which we now examine in turn.

\begin{enumerate}

\item ERBs are just the tail of the distribution of FRBs.  This seems the
least likely explanation.

\item ERBs arise from some systematic or instrumental effect, perhaps a local
household appliance masquerading as burst signals.  This kind of explanation
has no precedent and is thus also very unlikely.

\item ERBs are
related to a mild form of Lorentz violation.  All that is required is
that the signal in one special direction appears to be going faster than the
speed of light.  Surely no one would suggest such a thing to explain
puzzling experimental data.

\item This leaves the most plausible ERB explanation being
fabrication by hyper-intelligent aliens.  Conventional FRBs have already been
claimed to be of alien origin \cite{loeb1}.  Moreover, when the object
`Oumuamua passed through the Solar System there were suggestions that it might
be an artificial structure \cite{loeb2}, after which efforts were made
to look for SETI-like signals
from it \cite{oumuamua}.
The null results of these experiments are easily explained by the fact that
no one thought to look for {\it E\/}RBs that would appear before the search
even started.

\end{enumerate}

\section{Relationship to other types of burst}

Other kinds of transient event are also known.  Gamma-ray bursts (GRBs), were
first discovered in the 1960s and determined to be extragalactic explosions
after three decades of study.  They are now recognised to be the most luminous
electromagnetic events in the Universe.  Many models have been
proposed to explain GRBs \cite{grbs}, including that they may be
signals deliberately generated by aliens \cite{ball} or perhaps
extraterrestrial civilisations winking out of existence \cite{spider}.

Fast radio bursts (FRBs) are now more than a decade old \cite{decade}, and
their origin is still largely a mystery.  Nevertheless, aliens are again
being invoked as the explanation.

Now we also have a third class of source, the early-riser bursts (ERBs),
with non-Earth intelligence being their most likely origin.

How exactly are these different kinds of bursts related to each other
and to other kinds of transient source?
We would like to propose that whatever they turn out to be, the
next type of burst to be discovered will be the DRBs.  These will no doubt
be followed by CRBs, BRBs and ultimately by ARBs.  We would also suggest that
if old enough data are examined carefully, then there may well be evidence
for other types of transient event, such as HRBs, existing before the
discovery of GRBs.

\section{Conclusions}

Several recently discovered astronomical observations have their
explanations in alien activity, with ERBs being just the latest example.
With astronomers now embracing the idea that we are regularly seeing
phenomena coming from extraterrestrial intelligence \cite{sherlock}, we
should also be considering other creative explanations for mysterious
events in the sky.

The present authors would like to point to some related ideas of their own
that have already been published \cite{us}.  However, we feel that further
progress can only be made by being even more open-minded and imaginative.
There are already serious discussions in the literature regarding ideas
such as parallel universes, worm-holes, cosmic strings and variable speed
of light theories.  We believe that astronomers should be thinking about
combinations of such explanations for celestial phenomena.
For example, we
should be considering explanations for GRBs that involve extraterrestrial
civilizations {\it and\/} doomsday devices, FRB models that come from alien
megastructures {\it plus\/} warp drives, and ERBs having their origin
in ETs {\it using\/} time travel.  In fact, perhaps the best explanations
will ultimately come from multiple combinations, such as teleporting
aliens from alternative realities being seen through wormholes \cite{further}.


\smallskip



\begin{thebibliography}{9}
\bibitem{cade} OK, we made this one up.
\bibitem{ets} Some might see this as a trend for professional
astronomers to admit that they mostly wanted to study aliens rather than
astrophysics all along.
\bibitem{wick} To understand this, consider a system with cylindrical symmetry,
perhaps like a candle.  ``Wick rotations'' are those that just rotate the
central part, while leaving the outer part unrotated.
\end{thebibliography}
\end{document}